\documentclass[journal]{IEEEtran}
\usepackage{dsfont}
\usepackage{amssymb}
\usepackage{graphicx}
\usepackage[cmex10]{amsmath}
\usepackage{array}
\usepackage{epsfig}
\usepackage{color}
\usepackage[noadjust]{cite}
\usepackage{hyperref}
\hypersetup{
    colorlinks,%
    citecolor=black,%
    filecolor=black,%
    linkcolor=black,%
    urlcolor=black
}
\begin{document}
\title{Performance Analysis of Optimum Zero-Forcing Beamforming with Greedy User Selection}
\author{Serdar~Ozyurt and~Murat~Torlak,~\IEEEmembership{Senior Member,~IEEE}\\% <-this % stops a space
\thanks{The authors are with Electrical Engineering Department, University of Texas at Dallas, Richardson,
TX, 75080 USA (e-mail:serdarozyurt@utdallas.edu, torlak@utdallas.edu).}}
\markboth{IEEE Communications Letters, April~2012}%
{Shell \MakeLowercase{\textit{et al.}}: Bare Demo of IEEEtran.cls for Journals}
\newcommand {\SINR} {\textrm{SINR}}
\IEEEaftertitletext{\vspace{-3\baselineskip}}
\maketitle
\begin{abstract}
%\boldmath
In this letter, an exact performance analysis is presented on the sum rate of zero-forcing beamforming with a greedy user scheduling algorithm in a downlink system. Adopting water-filling power allocation, we derive a compact form for the joint probability density function of the scheduled users' squared subchannel gains when a transmitter with multiple antennas sends information to at most two scheduled users with each having a single antenna. The analysis is verified by numerical results.
\end{abstract}
\vspace{-.05in}
\begin{IEEEkeywords}
Zero-forcing beamforming, water-filling power allocation, user scheduling, sum rate.
\end{IEEEkeywords}
\IEEEpeerreviewmaketitle
%\vspace{-.15in}
\section{Introduction}
%\indent Use of multiple antennas at either side of any communication link has been attracted a lot of attention as it paves the way to obtain higher data rates without any increase in bandwidth or power consumption. Information theoretical results prove that dirty paper coding (DPC) is the capacity-achieving approach in a downlink multi-antenna channel where a base station (BS) having channel state information (CSI) available and equipped with $M$ antennas communicates $K$ single-antenna users [1]. Despite being the capacity-achieving technique, the implementation of DPC is still a challenging task with the current technology.
%In order to address user scheduling and transmit precoding problems jointly at relatively low complexity, some sub-optimal zero-forcing beamforming (ZFBF) methods combined with user scheduling have been proposed and shown to achieve the same asymptotic scaling as the capacity-achieving dirty paper coding (DPC) on the sum rate [1], [2]. In a ZFBF system, each user is assigned a beamforming (BF) weight vector that is within the null space of the other users' channels.
A greedy user selection algorithm combined with zero-forcing beamforming (ZFBF) has been proposed in~\cite{Dimic:05} as a lower complexity alternative to the capacity-achieving dirty paper coding (DPC) in a downlink transmission system. This scheme captures a significant fraction of the sum capacity. A similar combination of ZFBF and a greedy user selection algorithm has been shown to achieve full asymptotic DPC sum rate in~\cite{Wang:08}. Also, the authors propose a near-optimum method for moderate number of users by using LQ decomposition and pruning the search space. This scheme with a lower complexity as compared to the scheme in~\cite{Dimic:05} optimizes ZFBF weight vectors (called optimum ZFBF in this context) by using a composite channel matrix formed only from active users.

Most of the results on the sum rate performance analysis of ZFBF with user selection have been based on asymptotic analysis and/or the combination of ZFBF with DPC. However, the performance study for practical scenarios is important from an implementation point of view. A performance analysis on zero-forcing receivers has been presented in~\cite{Chen:07} where a base station (BS) with multiple antennas transmits information to a subset of users each employing ZFBF with multiple antennas. For scalar and vector feedback cases, the authors obtain analytical expressions for the sum rate. A sum rate analysis on a semi-orthogonal user selection algorithm combined with ZFBF at the BS is provided in~\cite{Lu:09} for two transmit antennas. Difficulty with ZFBF performance analysis is that the subchannel gains are the inverse of the diagonal elements of a matrix which has an inverted complex Wishart distribution. This joint probability density function (PDF) called multivariate gamma distribution has been studied for a $2\times 2$ matrix case in~\cite{Xu:09}. When some kind of user scheduling is applied, this result cannot be used directly due to ordering. With the greedy user scheduling~\cite{Dimic:05}, another difficulty comes from the fact that the subchannel gains are dependent on each other and a decision on any scheduling step directly affects the subchannel gains of previously scheduled users. In this letter, we provide a framework to study sum rate performance of ZFBF with a greedy user selection algorithm and water-filling power allocation (WPA) under average (long-term) power constraint at the transmitter, which does not exist in the literature to best of our knowledge. Assuming each user has a single antenna, we derive a joint PDF expression for the scheduled users' squared subchannel gains in closed-form when at most two users are scheduled. We derive cutoff value for WPA and verify our analysis with numerical results.

Notation: We use uppercase and lowercase boldface letters to represent matrices and vectors, respectively. The operators $\|.\|$, $(.)^{T}$, $(.)^{H}$, $|.|$, $\setminus$, and $\cap$ denote norm, transpose, Hermitian transpose, cardinality, set difference, and intersection of two sets, respectively.
%\vspace{-.15in}
\section{System Model and Optimum Zero-Forcing Beamforming with Greedy User Selection}
A downlink broadcast channel with $M$ antennas at BS and $K$ single-antenna users is considered with $\{K, M\} \geq 2$. The received signal at the $k$th user is modeled as $r_{k}=\textbf{h}_{k}^{T}\textbf{s}+e_{k}$ for $k \in \{1,\ldots,K\}$ where $\textbf{h}_{k}^{T}=\left[h_{k}(1)~h_{k}(2)~\ldots~h_{k}(M)\right]$ is the channel vector, $\textbf{s}\in \mathbb{C}^{M\times 1}$ is the transmitted signal, and $e_{k}\in \mathbb{C}$ is additive white Gaussian noise with zero mean and unit variance for $k$th user, respectively. We assume that channel vectors are independent and identically distributed (IID) and remain constant throughout one codeword transmission with independently changing between transmissions. We also assume a homogeneous network with enough spacings between the transmit antennas where the elements of $\textbf{h}_{k}$ are IID zero-mean complex Gaussian random variables with unit variance. The BS with perfect channel state information schedules two users at most and forms a BF weight vector for each scheduled user such that no interference is received at any scheduled user. Denoting the $i$th scheduled user by $k_i$ and the set of scheduled users by $U_{n}$ such that $k_{i} \in U_{n}\subset \{1,\ldots,K\}$ and $|U_{n}| \in \{1, 2\}$, the received signal at the $i$th scheduled user can be expressed as
%\vspace{-.015in}
\begin{equation}
r_{k_{i}}=\sqrt{P_i}\textbf{h}_{k_{i}}^{T}\textbf{w}_{k_{i}}x_{k_{i}}+\textbf{h}_{k_{i}}^{T}
\sum_{j\neq i}\sqrt{P_j}\textbf{w}_{k_{j}}x_{k_{j}}+e_{k_{i}}
%\vspace{-.05in}
\end{equation}
where $\textbf{w}_{k_{i}} \in \mathbb{C}^{M\times 1}$ is the unit-norm beamforming weight vector and $x_{k_{i}}\in \mathbb{C}$ is the unit-norm transmitted data symbol for $i$th scheduled user, respectively. Thus, the transmitted signal can be written as $\textbf{s}=\sum_{i}\textbf{w}_{k_i}x_{k_i}$. The transmitter is subject to an average power constraint such that $\sum_{i=1}^{|U_n|}\mathds{E}[P_i]\leq P$ with $P$ and $P_i$ representing available average transmit power and the allocated power for the $i$th scheduled user, respectively. The ZFBF matrix $\textbf{W}=[\textbf{w}_{k_{1}}~\ldots~\textbf{w}_{k_{|U_n|}}]$ is determined by the Moore-Penrose inverse of $\textbf{H}=[\textbf{h}_{k_{1}}~\ldots~\textbf{h}_{k_{|U_n|}}]^{T}$. The complexity of the channel inversion is reduced by utilizing LQ decomposition as $\textbf{H}=\textbf{L}\textbf{Q}$ where $\textbf{L}$ is a lower-triangular square matrix and $\textbf{Q}$ is a matrix with orthonormal rows such that $\textbf{Q}\textbf{Q}^H=\textbf{I}$~\cite{Wang:08}. Hence, we have $\textbf{W}=\textbf{Q}^{H}\textbf{L}^{-1}\textbf{D}$ where $\textbf{L}^{-1}=[\textbf{g}_1,\ldots,\textbf{g}_{|U_n|}]$ and $\textbf{D}$ is a diagonal matrix with its $i$th diagonal being equal to $1/\|\textbf{g}_{i}\|$. Consequently, the received signal at the $i$th scheduled user can be written as
%\vspace{-.015in}
\begin{equation}
r_{k_{i}}=(\sqrt{P_i}/\|\textbf{g}_{i}\|)x_{k_{i}}+e_{k_{i}} \mbox{~~~for~~~} i \in \{1,\ldots,|U_n|\}.
%\vspace{-.05in}
\end{equation}
%The optimal power allocation is obtained such that the ZFBF sum rate expression is maximized
%under the constraint $\sum_{i=1}^{|U_n|} \mathds{E}[P_i] \leq P$. The solution to this optimization problem is dubbed as water-filling
Power allocation is determined by WPA and given by\linebreak $P_i=(1/\mu-\|\textbf{g}_i\|^2)_+$. Note that $(.)_+$ refers to $\operatorname{max}(0,.)$ and $\mu$ is obtained from $\sum_{i=1}^{|U_n|}\mathds{E}[P_i] = P$. The resulting sum rate is given by
%\vspace{-.05in}
\begin{equation}
C(U_n)=\sum_{i=1}^{|U_n|}\left(\log\left(\frac{1}{\mu\|\textbf{g}_i\|^2}\right)\right)_{+} \mbox{~with~~} |U_{n}| \in \{1, 2\}.
\label{eq:sumrate}
%\vspace{-.05in}
\end{equation}
The cutoff value $\mu$ depends only on the statistics of $\|\textbf{g}_i\|^2$ rather than its instantaneous realization. This fact facilitates the performance analysis in the next section. A greedy user selection algorithm quite similar to the one in~\cite{Dimic:05} is employed to schedule two users at most. The first scheduled user is determined in such a way that it has the largest channel gain and $C(U_1)=\log(1+P_1 \|\textbf{h}_{k_1}\|^2)$ is set with $U_1={k_1}$ when $\|\textbf{h}_{k_1}\|^2 > \mu$. If we have $\|\textbf{h}_{k_1}\|^2 \leq \mu$ on the other hand, no user is scheduled during that particular channel realization.
%In the selection of the first user, the aim is to benefit from multiuser diversity gain as much as possible.
In the former case, a second user is selected from the set $\{1,\ldots,K\}\setminus U_1$ such that the sum rate expression in (\ref{eq:sumrate}) is maximized and the corresponding $U_2$ and $C(U_2)$ are obtained. If we have $1/\|\textbf{g}_2\|^{2} > \mu$ and $C(U_{2}) > C(U_{1})$, the algorithm schedules two users from $U_2$. Otherwise, one user from $U_{1}$ is scheduled. Finally, ZFBF matrix is obtained from the composite channel matrix of the scheduled (active) users~\cite{Wang:08}. The BF and user scheduling algorithm described above is called optimum ZFBF throughout the paper.
%\vspace{-.15in}
\section{Performance Analysis}
In this section, we present a framework from order statistics to obtain PDF expressions of the scheduled users' squared subchannel gains.
%Let $y_{ij}$ denote $i$th scheduled user's squared subchannel gain when the number of scheduled users is $j$ with $1\leq i \leq j \leq 2$.
We start by introducing squared subchannel gains under random user scheduling (no user ordering) as $\{v_{k},z_{k}\}=\left\{\|\textbf{h}_{k}\|^2, \textbf{h}_{k}^H\textbf{P}_{1}^\perp\textbf{h}_{k}\right\}$ for $k \in \{1,2,\ldots,K\}$
%\begin{displaymath}
%\{v_{k},z_{k}\}=\left\{
%\begin{array}{lcr}
%v_1=\|\textbf{h}_{1}\|^2, ~v_1 \frac{z_2}{v_2}=\textbf{h}_{1}^H\textbf{P}_{\textbf{h}_{2}}^\perp\textbf{h}_{1}&\text{\!for~~$k=1$},\\
%v_k=\|\textbf{h}_{k}\|^2, ~~~z_k=\textbf{h}_{k}^H\textbf{P}_{\textbf{h}_{1}}^\perp\textbf{h}_{k}&\!\text{\!for~~$2\! \leq\! k\! \leq\! K$}
%       \end{array}\right.
%       \label{eq:equnorderedSINRs}
%       \end{displaymath}
where $v_{k}$ represents the resulting squared subchannel gain when the $k$th user is randomly scheduled at the first step. Similarly, $z_k$ denotes the resulting squared subchannel gain when the $k$th user is randomly scheduled at the second step (assuming it is not scheduled at the first step). Also, $\textbf{P}_{1}^\perp$ denotes a $(M-1)$ dimensional orthogonal projection matrix on the null space of the first scheduled user's channel vector. It is worth to mention that $\textbf{P}_{1}^\perp$ is independent from the decision at the first step for both random user scheduling and the greedy scheduling algorithm described above~\cite{Dimic:05}. The variables $\{v_{k},z_{k}\}$ have chi-squared distributions with $2M$ and $2(M-1)$ degrees of freedom, respectively. Their joint PDF is given by
%\vspace{-.05in}
\begin{equation}
f_{V,Z}(v_{k},z_{k})=\frac{z_{k}^{M-2}}{\Gamma(M-1)}~e^{-v_{k}} \mbox{~~for~~} ~~ v_{k} \geq z_{k} \geq 0.
\label{eq:jointpdf}
%\vspace{-.05in}
\end{equation}
Likewise, the composite joint PDF of $\{v_{1},z_{1},\ldots,v_{K},z_{K}\}$ can be written as
%\vspace{-.1in}
\begin{equation}
f_{V,Z}^{(c)}(\{v_{k},z_{k}\}_{k=1}^{K})=\prod_{k=1}^{K} f_{V,Z}(v_{k},z_{k}).
%e^{-\left(x_{1}+\ldots+x_{r}\right)}\frac{x_{r}^{M-r}}{\Gamma(M-r+1)}
\label{eq:unorderedjointpdf}
%\vspace{-.05in}
\end{equation}
%\vspace{-.135in}
%\begin{eqnarray}
%f(v_1)=\frac{v_{1}^{M-1}}{\Gamma(M)}~e^{-v_1} &\mbox{ for }& v_1 \geq 0 \mbox{~~~and~}\nonumber\\
%f(v_k,z_k)=\frac{z_{k}^{M-2}}{\Gamma(M-1)}~e^{-v_k} &\mbox{ for } & v_k \geq z_k \geq 0.
%%e^{-\left(x_{1}+\ldots+x_{r}\right)}\frac{x_{r}^{M-r}}{\Gamma(M-r+1)}
%\label{eq:unorderedjointpdf}
%\end{eqnarray}
Let us now apply the user scheduling and assume that it yields $\|\textbf{h}_{1}\|^{2}$ ($k_1 = 1$) and $\textbf{h}_{2}^H\textbf{P}_{\textbf{h}_{1}}^\perp\textbf{h}_{2}$ ($k_2 = 2$) for the first and second scheduling steps, respectively. Defining $\gamma_1=\|\textbf{h}_{1}\|^2$, \linebreak $\beta_1=\textbf{h}_{1}^H\textbf{P}_{\textbf{h}_{2}}^\perp\textbf{h}_{1}$, $\gamma_2=\|\textbf{h}_{2}\|^2$, and $\beta_2=\textbf{h}_{2}^H\textbf{P}_{\textbf{h}_{1}}^\perp\textbf{h}_{2}$, we can write $\beta_1=(\gamma_{1}/\gamma_{2})\beta_2$ with%
%
%
%
% $v_{11}=\|\textbf{h}_{1}\|^{2}=\underset{k}{\max}~v_{k1}$ and $v_{22}=\textbf{h}_{2}^H\textbf{P}_{\textbf{h}_{1}}^\perp\textbf{h}_{2}=\underset{k \geq 2}{\max}~v_{k2}$. Using $v_{12}=v_{11}(v_{22}/v_{21})$, the average sum rate can be written as
%\vspace{-.015in}
\begin{equation}
C(U_1)\!\!=\!\!\log\!\left(\frac{\gamma_{1}}{\mu}\right)\!\mbox{~and~}\hspace{.0005in}
~C(U_2)\!\!=\!\!\log\!\left(\frac{\gamma_{1}}
{\gamma_{2}}\frac{\beta_{2}^{2}}{\mu^2}\right)
%\mathds{E}\left[\log\left(\frac{v_1}{\mu}\frac{z_2}{v_2}\right)+\log\left(\frac{z_2}{\mu}\right)\right]=
\label{eq:CU1_CU2}
%\vspace{-.015in}
\end{equation}
when the number of the scheduled users with nonzero power allocations is one and two, respectively. The joint PDF of $\{\gamma_{1}, \gamma_{2}, \beta_{2} \}$ is required in order to determine the average sum rate. The following theorem provides a compact way to compute this joint PDF.
\newtheorem{theorem}{Theorem}
\begin{theorem}
The joint PDF of $\{\gamma_{1}, \gamma_{2}, \beta_{2}\}$ is given by
\begin{displaymath}
f_{\gamma_{1}, \gamma_{2}, \beta_{2}}(\gamma_{1}, \gamma_{2}, \beta_{2})=\frac{K!}{(K-2)!}e^{-(\gamma_{1}+\gamma_{2})}\gamma_{1}^{M-1}\beta_{2}^{M-2}
\end{displaymath}
\begin{displaymath}
\times \frac{M-1}{\Gamma(M)^K} \Bigg\{\Gamma(M)-\Gamma\left(M,\frac{\beta_{2}^{2}}{\gamma_{2}}\right)
+\left(\frac{\beta_{2}}{\sqrt{\gamma_{2}}}\right)^{M-1}
\end{displaymath}
\begin{equation}
\times\left(\Gamma\left(\frac{M+1}{2},\frac{\beta_{2}^{2}}{\gamma_{2}}\right)-\Gamma\left
(\frac{M+1}{2},\gamma_{1}\right)\right)\Bigg\}^{\!\!K-2}
\label{eq:teorem}
\end{equation}
for $\gamma_{1} \geq \gamma_2 \geq \beta_{2} \geq 0$. In (\ref{eq:teorem}), $\Gamma(s)$ and $\Gamma(s, x)$ respectively denote the gamma and upper incomplete gamma functions~\cite{Gradshteyn:00}.
\end{theorem}
\begin{IEEEproof}
Considering the greedy user selection procedure described in the previous section, two conditions namely $\gamma_{1} \geq \{v_{2},\ldots,v_{K}\}$ and $\beta_{2}^{2}/\gamma_{2}\geq\{z_{3}^{2}/v_{3},\ldots,z_{K}^{2}/v_{K}\}$ must be satisfied, which correspond to the first two scheduling steps, respectively. Note that since $\gamma_{1}/\mu^2$ is common for all candidate users, it has no effect on the scheduling at the second step. Additionally, $K!/(K-2)!$ different 2-permutations can be selected out of $K$ users as the first two scheduled users. Consequently, resorting to order statistics~\cite{David:03} and using (\ref{eq:unorderedjointpdf}), the joint PDF of $\{\gamma_{1}, \gamma_{2}, \beta_{2} \}$ can be expressed as
\begin{equation}
f_{\gamma_{1}, \gamma_{2}, \beta_{2}}(\gamma_{1}, \gamma_{2}, \beta_{2})\!=\!\frac{K!}{(K-2)!}\left(\int_{0}^{\gamma_{1}}\!\!f_{V,Z}(\gamma_{1},z_{1})dz_{1}\right)
\label{eq:jointpdf1}
\end{equation}
\begin{displaymath}
\times f_{V,Z}(\gamma_{2},\beta_{2})\! \int_{S}\!\left(\prod_{k=3}^{K}f_{V,Z}(v_k,z_k)\!\right)dv_{3}dz_{3}\ldots dv_{K}dz_{K}
%\label{eq:jointpdf1}
\end{displaymath}
with $S\!=\!\!\Big\{v_{3},z_{3},\!\ldots\!,v_{K}, z_{K}\!:\!\{z_{3}\! \leq \! v_{3}\! \leq \!\gamma_{1},\!\ldots\!,z_{K}\! \leq \! v_{K}\! \leq \!\gamma_{1} \}$\linebreak
$
\cap~\{z_{3}^{2}/v_{3} \leq \beta_{2}^{2}/\gamma_{2},\ldots,z_{K}^{2}/v_{K} \leq \beta_{2}^{2}/\gamma_{2}\}\Big\}.
$
Using $S$, (\ref{eq:jointpdf1}) can be written as
\begin{displaymath}
f_{\gamma_{1}, \gamma_{2}, \beta_{2}}(\gamma_{1}, \gamma_{2},\beta_{2})\!=\!\frac{K!}{(K-2)!}\left(\int_{0}^{\gamma_{1}}\!\!f_{V,Z}(\gamma_{1},z_{1})dz_{1}\right)
\end{displaymath}
\begin{equation}
\times f_{V,Z}(\gamma_{2},\beta_{2}) \left(\int_{S_1}\!\!\!f_{V,Z}
(v_{3},z_{3})dv_{3}dz_{3}\right)^{K-2}
\end{equation}
with $S_1=\left\{v_{3},z_{3}:\{z_{3}\leq v_{3}\leq \gamma_{1}\}\cap \{z_{3}^{2}/v_{3}\leq \beta_{2}^{2}/\gamma_{2}\}\right\}$. Properly dividing $S_1$ into two disjoint regions and using (\ref{eq:jointpdf}), we can write
\begin{displaymath}
f_{\gamma_{1}, \gamma_{2}, \beta_{2}}(\gamma_{1},\gamma_{2},\beta_{2})=\frac{K!}{(K-2)!}
\frac{e^{-(\gamma_{1}+\gamma_{2})} ~\gamma_{1}^{M-1}~ \beta_{2}^{M-2}}{\Gamma(M)~ \Gamma(M-1)^{K-1}}
\end{displaymath}
\begin{equation}
\!\!\times\!\!\left(\!\!\int_{0}^{\frac{\beta_{2}^{2}}{\gamma_{2}}}\!\!\!\!\int_{0}
^{v_{3}}\!\!\!\!\!\!e^{-v_{3}}z_{3}^{M-2}dz_{3} dv_{3}\!+\!\!\int_{\frac{\beta_{2}^{2}}{\gamma_{2}}}^{\gamma_{1}}\!\!\int_{0}^{\sqrt{\frac{\beta_{2}^{2}v_{3}}{\gamma_{2}}}}\!\!
\!\!\!\!e^{-v_{3}}z_{3}^{M-2}dz_{3} dv_{3}\!\!\right)^{\!\!\!\!K-2}\!\!.
\end{equation}
Evaluation of the preceding integrals leads to the result in (\ref{eq:teorem}).
\end{IEEEproof}
\begin{figure}[!t]
\vspace{-5mm}
\advance\leftskip-.15in
\includegraphics[width=0.55\textwidth]{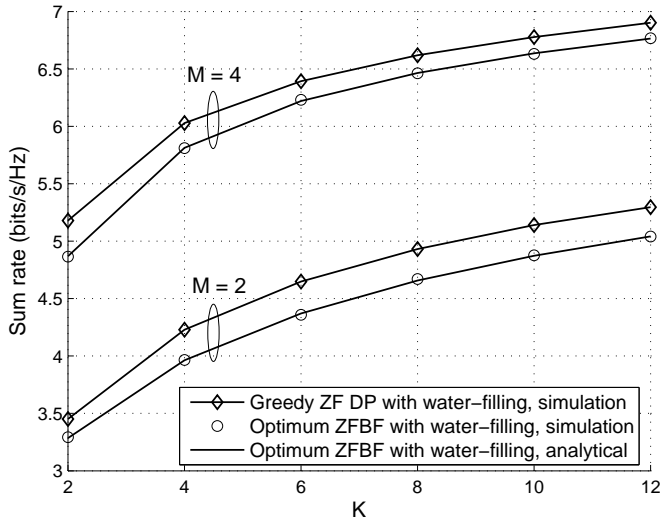}
\vspace{-8mm}
\caption{Analytical and simulated sum rate results on optimum ZFBF together with that of greedy ZF DP for $P=5$ dB, $M=\{2, 4\}$, and varying $K$.}
%1e4realizations kullanildi.
\vspace{-1mm}
%\vspace{-.15in}
\end{figure}
%\vspace{-.15in}
\subsection{Derivation of Cutoff Value}
The joint PDF derived above can be used to determine the cutoff value $\mu$ from the power constraint. Only one user is scheduled with an allocated power of $(1/\mu-1/\gamma_{1})$ if one of two following conditions holds: $\{\beta_{2} \geq \mu \cap \frac{\gamma_{1} \beta_{2}^2}{\gamma_{2} \mu^2}\leq \frac{\gamma_{1}}{\mu}\}$ and $\{\beta_{2} \leq \mu \cap \gamma_{1} \geq \mu\}$. Similarly, the condition is given by $\{\beta_{2} \geq \mu \cap \frac{\gamma_{1} \beta_{2}^2}{\gamma_{2} \mu^2}\geq \frac{\gamma_{1}}{\mu}\}$ for the case of two scheduled users with a total assigned power of $(2/\mu-\gamma_{2}/ (\gamma_{1} \beta_{2})-1/\beta_{2})$. Consequently, the power constraint can be expressed as the average of the allocated sum power values over these three disjoint regions as follows
%\vspace{-.05in}
\begin{displaymath}
P=\int\limits_{\mu}^{\infty}\int\limits_{\mu}^{\sqrt{\mu \gamma_{1}}}\int\limits_{\frac{\beta_{2}^{2}}{\mu}}^{\gamma_{1}}\left(\frac{1}{\mu}-\frac{1}{\gamma_{1}}\right)f_{\gamma_{1}, \gamma_{2}, \beta_{2}}(\gamma_{1},\gamma_{2},\beta_{2})d\gamma_{2} d\beta_{2} d\gamma_{1}
\end{displaymath}
%\vspace{-.05in}
\begin{equation}
+\int\limits_{\mu}^{\infty}\int\limits_{0}^{\mu}\int\limits_{\beta_{2}}^{\gamma_{1}}\left(\frac{1}{\mu}-\frac{1}{\gamma_{1}}\right)f_{\gamma_{1}, \gamma_{2}, \beta_{2}}(\gamma_{1},\gamma_{2},\beta_{2})d\gamma_{2} d\beta_{2} d\gamma_{1}
\label{eq:cutoff}
\end{equation}
%\vspace{-.05in}
\begin{displaymath}
+\int\limits_{\mu}^{\infty}\int\limits_{\mu}^{\sqrt{\mu \gamma_{1}}}\int\limits_{\beta_{2}}^{\frac{\beta_{2}^{2}}{\mu}}\left(\frac{2}{\mu}-\frac{\gamma_{1}+\gamma_{2}}{\gamma_{1} \beta_{2}}\right)f_{\gamma_{1}, \gamma_{2}, \beta_{2}}(\gamma_{1},\gamma_{2},\beta_{2})d\gamma_{2} d\beta_{2} d\gamma_{1}
\end{displaymath}
%\vspace{-.05in}
\begin{displaymath}
+\int\limits_{\mu}^{\infty}\!\int\limits_{\sqrt{\mu \gamma_{1}}}^{\gamma_{1}}\int\limits_{\beta_{2}}^{\gamma_{1}}\!\left(\frac{2}{\mu}-\frac{\gamma_{1}+\gamma_{2}}{\gamma_{1} \beta_{2}}\right)f_{\gamma_{1}, \gamma_{2}, \beta_{2}}(\gamma_{1},\gamma_{2},\beta_{2}) d\gamma_{2} d\beta_{2} d\gamma_{1}.
\end{displaymath}
The cutoff value $\mu$ can be calculated for a given set of $\{K,M,P\}$ values from (\ref{eq:cutoff}) using standard mathematical software packages. Note that when the power allocation values are properly replaced by $C(U_1)$ and $C(U_2)$ given in (\ref{eq:CU1_CU2}), the right-hand side of (\ref{eq:cutoff}) yields the average sum rate result.
\section{Numerical Results}
Greedy zero-forcing dirty-paper coding (ZF DP) algorithm is a combination of scalar DPC, beamforming, and greedy user scheduling at the transmitter and shown to attain same slope of sum rate increase with transmit power in dB as the capacity-achieving DPC~\cite{Dimic:05}. In Fig. 1, we illustrate the simulated and analytically obtained sum rate results on optimum ZFBF for $M=\{2,4\}$, $P=5$ dB, and varying $K$. The performance of greedy ZF DP with WPA under a long-term power constraint is also plotted as a benchmark. The strong match between the simulated and analytical results verifies the accuracy of our analytical derivation. Additionally, optimum ZFBF yields a sum rate that is more than $90\%$ of the sum rate of greedy ZF DP for both cases with a much lower complexity~\cite{Dimic:05}. For both optimum ZFBF and greedy ZF DP, inverse of the cutoff value is plotted in Fig. 2 with $M=\{2, 4\}$, $P=\{0, 5\}$ dB, and varying $K$.
%The inverse of cutoff value
%decreases as $K$ increases.
When $P$ is small, $\mu$ value is set higher as expected and growth rate of $\mu$ in $K$ slows down as $K$ increases.
% and the dependence of $\mu$ appears to be logarithmic on $K$.
%The offered gain of adaptive user selection and water-filling can be seen clearly for small $K$ values. As $K$ increases, the channels of the users behave much like uniformly and the gain against uniform power allocation diminishes. Fig. 2 is plotted for $M=\{2,4\}$, $K=3$, and varying $P$ where the simulated and analytical sum rate expressions are depicted. When $M=4$, water-filling scheme outperforms uniform power allocation for small $P$ values and they have similar performances otherwise. Also, water-filling algorithm performs better for all plotted $P$ values when $M=2$. Above all, analytical and simulated results are on top of each other for both figures clearly corroborating the accuracy of the analytical derivation in the previous section.
%\begin{figure}[!t]
%\vspace{-5mm}
%\advance\leftskip-.15in
%\includegraphics[width=0.55\textwidth]{ZFS_Fig1}
%\vspace{-5mm}
%\caption{Comparison of water-filling with uniform power allocation for $M=\{2,4\}$, $P=5$ dB, and varying $K$.}
%%1e4realizations kullanildi.
%%\vspace{-.1mm}
%%\vspace{-.15in}
%\end{figure}
\begin{figure}[!t]
\vspace{-5mm}
\advance\leftskip-.15in
\includegraphics[width=0.55\textwidth]{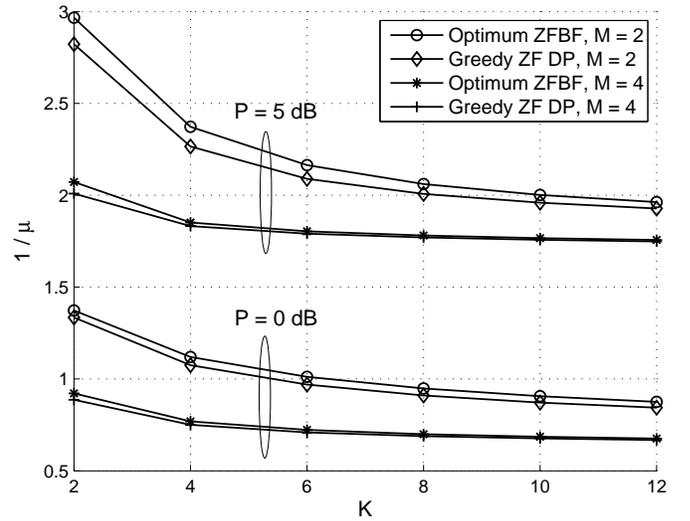}
\vspace{-8mm}
\caption{Inverse of the cutoff value for optimum ZFBF and greedy ZF DP with $M=\{2, 4\}$, $P=\{0, 5\}$ dB, and varying $K$.}
%1e4realizations kullanildi.
\vspace{-1mm}
%\vspace{-.1in}
\end{figure}
%\vspace{-.15in}
\section{Conclusion}
In this paper, we present a sum rate performance analysis on zero-forcing beamforming combined with a greedy user scheduling algorithm in a downlink system with single-antenna users. Assuming the number of scheduled users is at most two, a mathematical framework has been introduced including water-filling power allocation under an average power constraint. Using this framework, we have found a closed-form expression for the joint probability density function of the scheduled users' squared subchannel gains. The framework has enabled us to express the cutoff value for water-filling power allocation in a suitable way to solve. The analysis has been numerically verified. %The presented framework will be used to perform the analysis for more than two scheduled users in a forthcoming publication.
%\vspace{-.15in}

\end{document}